\documentclass[sigconf,screen,authorversion,nonacm]{acmart}
\AtBeginDocument{%
  }

\copyrightyear{2025}
\acmYear{2025}
\setcopyright{cc}
\setcctype{by-nc-nd}
\acmConference[ICCSIT 2025]{The 18th International Conference on Computer Science and Information Technology}{October 27--29, 2025}{Paris, France}
\acmBooktitle{The 18th International Conference on Computer Science and Information Technology (ICCSIT 2025), October 27--29, 2025, Paris, France}
\acmPrice{}
\acmDOI{10.1145/3783862.3783877}
\acmISBN{979-8-4007-1858-8/2025/10}

\begin{document}

\title{Waterfall Model Simulation: A Systematic Mapping Study}

\author{Antonios Saravanos}
\email{saravanos@nyu.edu}
\orcid{0000-0002-6745-810X}
\affiliation{%
  \institution{New York University}
  \streetaddress{7 East 12th Street, Room 625B}
  \city{New York}
  \state{NY}
  \country{USA}
  \postcode{10003}
}

\renewcommand{\shortauthors}{Saravanos, A.}

\begin{abstract}
This paper systematically maps peer-reviewed research and graduate theses/dissertations that explicitly simulate the waterfall model. Following Petersen’s mapping guidelines and Kitchenham’s systematic literature review practices, major databases (ACM Digital Library, IEEE Xplore, Scopus, Springer, Google Scholar, and Web of Science) were searched for studies published between 2000 and 2024 using the title query (``simulation'' OR ``simulating'') AND ``waterfall''. A PRISMA workflow guided the screening process, and approximately 9\% of retrieved records met the inclusion criteria. A repeated extraction process captured methods, tools, venues, geography, publication years, comparative scope, and fidelity to Royce’s original model; findings were synthesized thematically. Discrete-event simulation dominates (80\%) compared to system dynamics (20\%). Reported tools center on Simphony.NET (40\%) and SimPy (20\%), while 40\% of studies omit tool details, limiting reproducibility. Research is distributed across Italy, Lebanon, India, Japan, and the United States; publication venues include 60\% journals and 40\% conferences. Sixty percent of studies are comparative, while 40\% model only the waterfall approach. No study reproduces Royce’s original model; all employ adaptations. The paper concludes by presenting a consolidated view of waterfall simulation research and recommending clearer model reporting, fuller tool disclosure, and wider adoption of open-source platforms.
\end{abstract}

\ccsdesc[500]{Software and its engineering~Software creation and management}
\ccsdesc[500]{Software and its engineering~Software development process management}

\keywords{waterfall model; software development lifecycle; systems development lifecycle; SDLC; simulation; systematic mapping study
}

\maketitle

\section{Introduction and Background}

\begin{figure*}
    \centering
    \includegraphics[width=1\linewidth]{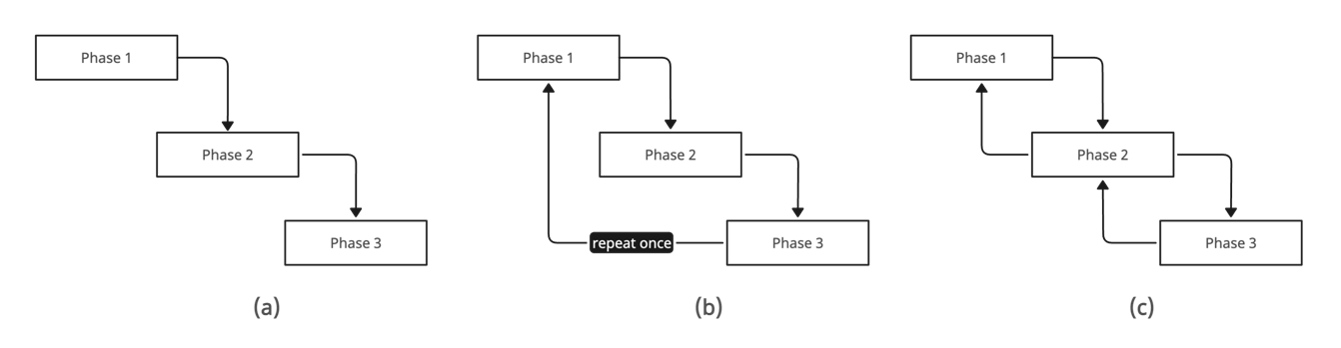}
     \caption{Three formulations of the waterfall model (adapted from Royce~\cite{royce1987}): (a) single pass with strictly sequential phases and minimal feedback; (b) two-pass formulation in which the full lifecycle is executed twice; and (c) single pass with explicit feedback loops to the immediately preceding phase(s).}
    \label{fig:waterfall}
\end{figure*}

Software development methodologies define how systems are planned, built, and maintained. Among these, the waterfall model remains one of the most historically influential, shaping decades of engineering practice and pedagogical design. Introduced by Royce~\cite{royce1987}, the waterfall model formalized a sequential approach to software development that progresses through fixed stages such as requirements analysis, design, implementation, testing, and maintenance. In practice, these phases are often combined or abbreviated, yet the core principle of irreversible, step-by-step progression remains. The model is characterized by its linear, sequential structure for developing systems and software solutions, as illustrated in Figure~\ref{fig:waterfall}. Although frequently presented as a simple linear framework (a), Royce described two other formulations: one that executes the lifecycle twice (b) and another that introduces feedback loops to earlier phases. Despite these differences in feedback mechanisms, all three are commonly grouped under the label ``waterfall''. Over time, the model has been widely adapted and variably interpreted across academic, industrial, and regulatory contexts. 

While research in recent years has primarily emphasized agile and iterative methods, there has been relatively little sustained investigation into how traditional models like waterfall are analyzed, taught, or critiqued through computational simulation. Simulation provides a powerful means of exploring software process behavior, enabling controlled experimentation with task durations, resource allocations, phase dependencies, and defect or failure scenarios without the risks of real-world implementation. It complements static lifecycle diagrams and serves as a valuable instructional tool for demonstrating the consequences of early-stage decisions, structural rigidity, and process bottlenecks. In this paper, ``simulation'' means a computer-run model that plays out a process over time or events (e.g., discrete-event, system dynamics, agent-based, Monte Carlo). Non-computer activities, such as walkthroughs, reenactments, tabletop drills, role-play, mock demos, or spreadsheet what-ifs, are considered scenario-based demonstrations and are not considered in our study.

Despite these advantages, no comprehensive mapping study currently exists that systematically examines how the waterfall model has been represented and studied through simulation. Existing surveys of software process simulation tend to cover broader paradigms or concentrate on agile and hybrid practices, leaving a clear gap in the literature. Understanding how, why, and to what extent the waterfall model has been simulated can help clarify its continued relevance, reveal methodological trends, and guide the design of future simulation tools. This paper addresses this gap by conducting a systematic mapping study of simulation-based research published from 2000 to 2024 that explicitly simulates the waterfall model. By synthesizing this body of work, we aim to clarify the current state of simulation research on the waterfall model and establish a foundation for more rigorous, accessible, and context-aware simulation efforts. In doing so, we position simulation not only as a mode of academic analysis but also as a practical tool for education, planning, and process experimentation in software engineering.

Yet, despite its suitability for simulation, no systematic mapping study has focused exclusively on how the waterfall model has been simulated in the academic literature. Existing reviews tend to address software process simulation more broadly or concentrate on contemporary methodologies, often neglecting traditional models or mentioning them only in comparison. For instance, a recent systematic review by García-García et al.~\cite{garcia2020} on software process simulation identified 8{,}070 relevant publications from 2013 to 2019. However, that review examined software processes in general, and the term ``waterfall'' does not appear. 

This research addresses that gap by conducting a systematic mapping study of peer-reviewed work that explicitly simulates the waterfall model. Drawing on major academic databases and guided by established protocols for systematic mapping in software engineering, this study synthesizes two decades of scholarship in order to: (1) identify and classify the simulation methodologies used to model the waterfall process; (2) examine the software tools and platforms employed in these simulations; (3) assess the fidelity, transparency, and reproducibility of the models presented; (4) trace the evolution of scholarly interest across geographies and publication venues; and (5) highlight gaps, limitations, and opportunities for future research. By consolidating this fragmented body of work, the study contributes to a clearer understanding of the current state of research on waterfall model simulation.

The remainder of this paper is organized as follows. Section~2 details the methodology used for conducting the systematic mapping study, including the research questions, search strategy, and inclusion criteria. Section~3 presents the results, focusing on trends in simulation methodologies, tools, geographic and temporal patterns, comparative scope, and fidelity to the original waterfall model. Section~4 discusses limitations and opportunities for future research. Section~5 concludes with a summary of key findings and their implications.

\section{Materials and Methods}

This section outlines the approach taken for this systematic mapping study, following the guidelines of Petersen et al.~\cite{petersen2008} to ensure that the process is transparent, rigorous, and replicable. For generic procedures (e.g., search construction, screening workflow, and data-extraction forms), we adapted established systematic literature review practices from Kitchenham et al.~\cite{kitchenham2010}. An adapted PRISMA 2020 flow diagram (identification, screening, eligibility, inclusion) is included for transparency~\cite{page2021}. Consistent with the aims of systematic mapping studies, we did not perform an appraisal of evidence strength (cf. Dybå \& Dingsøyr~\cite{dyba2008}) and conducted no meta-analysis.

\subsection{Research Objectives and Questions}

The objective of this study is to provide a structured and comprehensive overview of how the waterfall model has been represented through simulation in the academic literature. To achieve this, the study formulates a set of research questions that serve as analytical lenses for categorizing, synthesizing, and interpreting the findings. The rationale for including these research questions is twofold. First, they allow for a systematic partitioning of the literature along dimensions that are both theoretically meaningful and practically relevant, such as simulation methods, tools, and domains of application. Second, they are intended to surface underexplored areas and highlight emerging patterns or inconsistencies that may guide future research and practice.

To this end, the research questions were organized into four thematic categories. The first theme focuses on simulation methodologies and tools, aiming to identify the dominant technical approaches used to represent the waterfall model (RQ1.1) and the software platforms that support such representations (RQ1.2). This category is essential for understanding methodological diversity and the field’s technical evolution. The second theme addresses geographic contributions (RQ2.1), mapping where research activity is concentrated and identifying potential regional trends or disparities. The third theme concerns publication trends (RQ3.1--RQ3.3), examining the types of venues, the extent to which simulations focus exclusively on the waterfall model versus comparative contexts, and the timeline of research activity. This category provides insight into the maturity of the field and shifting scholarly interest in simulating traditional software processes. The fourth theme evaluates model fidelity (RQ4.1), asking whether studies adhere to Royce’s original waterfall model or employ adaptations. This question is particularly relevant given the diversity of interpretations of the model in both academic and professional contexts. Together, these research questions form the foundation for data extraction and thematic analysis and are listed in full in Table~\ref{tab:research-questions}.

\begin{table*}[htbp]
\caption{Summary of Research Questions}
\label{tab:research-questions}
\centering
\renewcommand{\arraystretch}{1.25}
\begin{tabular}{p{3.2cm} p{6.2cm} p{5.8cm}}
\toprule
\textbf{Theme} & \textbf{Research Question} & \textbf{Purpose} \\
\midrule
Methodologies and Tools &
RQ1.1: What methodologies are used to simulate the waterfall model? &
To identify the modeling approaches employed in waterfall simulations (e.g., discrete-event simulation, system dynamics). \\

&
RQ1.2: What tools or software are employed in simulations of the waterfall model? &
To identify the technical platforms or frameworks commonly used in simulation studies. \\
\midrule
Geographic Contributions &
RQ2.1: What are the geographic locations of authors contributing to this research domain? &
To examine the regional distribution of scholarly activity in waterfall simulation research. \\
\midrule
Publication Trends &
RQ3.1: In what types of publications does this research appear (e.g., journals, conferences)? &
To assess the academic venues and dissemination patterns of relevant studies. \\

&
RQ3.2: Do the studies focus exclusively on the waterfall model, or do they include comparisons with others? &
To determine the scope of simulation efforts (waterfall-only versus comparative analyses). \\

&
RQ3.3: In which years were studies on simulating the waterfall model published? &
To trace the chronological evolution and intensity of research interest in this topic. \\
\midrule
Model Variants &
RQ4.1: Was Royce’s original waterfall model simulated, or was a variation used? &
To examine model fidelity and the extent to which simulations adhere to the original formulation. \\
\bottomrule
\end{tabular}
\end{table*}

\subsection{Search Strategy}

To identify relevant literature, a structured search strategy was developed and applied across a range of leading academic databases, listed in Table~\ref{tab:databases}. The search was designed to maximize sensitivity and recall through broad and inclusive search terms. The search string used was: (``simulation'' OR ``simulating'') AND ``waterfall''. This query was applied to the title field of indexed articles to ensure that the simulation of the waterfall model was a primary focus of the study. All searches were restricted to peer-reviewed articles published in English between January 1, 2000, and December 31, 2024, inclusive. Boolean operators were employed where supported. The term ``waterfall'' was used on its own, without combinations such as ``waterfall model'' or ``SDLC'', to avoid inadvertently excluding relevant papers. The databases were last checked on June 25, 2025. While some databases, such as Scopus, supported advanced search functionalities, others, including Google Scholar, required manual filtering due to more limited search controls.

\begin{table}[ht]
    \centering
    \caption{Selected Databases}
    \label{tab:databases}
    \renewcommand{\arraystretch}{1.2}
    \begin{tabular}{ll}
        \toprule
        \textbf{Source} & \textbf{Location} \\
        \midrule
        ACM Digital Library & \url{https://dl.acm.org} \\
        IEEE Xplore & \url{https://ieeexplore.ieee.org} \\
        Scopus & \url{https://www.scopus.com} \\
        Springer & \url{https://link.springer.com} \\
        Google Scholar & \url{https://scholar.google.com} \\
        Web of Science & \url{https://www.webofscience.com} \\
        \bottomrule
    \end{tabular}
\end{table}

\subsection{Inclusion and Exclusion Criteria}

To ensure methodological rigor and relevance, we define simulation as a computer-run model that plays out a process over time or events (e.g., discrete-event, system dynamics, agent-based, Monte Carlo), and we do not consider scenario-based demonstrations such as walkthroughs, reenactments, tabletop drills, role-play, mock demos, or spreadsheet what-ifs in our study. Eligible studies included peer-reviewed journal articles, conference proceedings, and graduate theses or dissertations (master’s or doctoral) that explicitly simulated the waterfall model, either on its own or in comparison with other models. Each study had to provide sufficient context to interpret the simulation (e.g., objectives, model description, assumptions, and data or parameters, where applicable) and be available in full text. Only English-language publications dated between January 1, 2000, and December 31, 2024, were considered. Studies were excluded if they consisted of non-scholarly gray literature (e.g., technical reports without academic review, blog posts), used the term ``waterfall'' in an unrelated context (e.g., geological phenomena), did not perform a simulation, lacked methodological transparency, or were unavailable in full text. These criteria were applied at both the abstract and full-text screening stages, with database filtering tools used when available. Table~\ref{tab:inclusion_exclusion} summarizes the inclusion and exclusion criteria.

\begin{table*}[htbp]
\caption{Summary of Inclusion and Exclusion Criteria}
\label{tab:inclusion_exclusion}
\centering
\renewcommand{\arraystretch}{1.2}
\begin{tabular}{p{3.2cm} p{5.8cm} p{5.8cm}}
\toprule
\textbf{Criterion Type} & \textbf{Inclusion Criteria} & \textbf{Exclusion Criteria} \\
\midrule
Publication Type &
Peer-reviewed journal articles, conference proceedings, and doctoral or master’s theses/dissertations. &
Editorials, opinion pieces, blog posts, or non-peer-reviewed literature. \\

Language &
Articles published in English. &
Articles published in languages other than English. \\

Publication Year &
Studies published between January~1,~2000, and December~31,~2024. &
Studies published outside this date range. \\

Relevance of Content &
Explicit computational simulation of the waterfall model, either standalone or comparative. ``Simulation'' refers to a computer-executed model. &
Studies using ``waterfall'' in unrelated contexts (e.g., geological) or lacking computational simulation; scenario-based demonstrations such as walkthroughs, reenactments, tabletop drills, role-play exercises, mock demonstrations, or spreadsheet-based what-if analyses. \\

Methodological Rigor &
Clear research objectives, detailed simulation methodology, and sufficient contextual information. &
Studies lacking methodological transparency or adequate simulation detail. \\

Accessibility &
Full text available for analysis. &
Studies unavailable in full text. \\
\bottomrule
\end{tabular}
\end{table*}

\subsection{Study Selection Procedure}

\begin{figure*}
    \centering
    \includegraphics[width=0.70\linewidth]{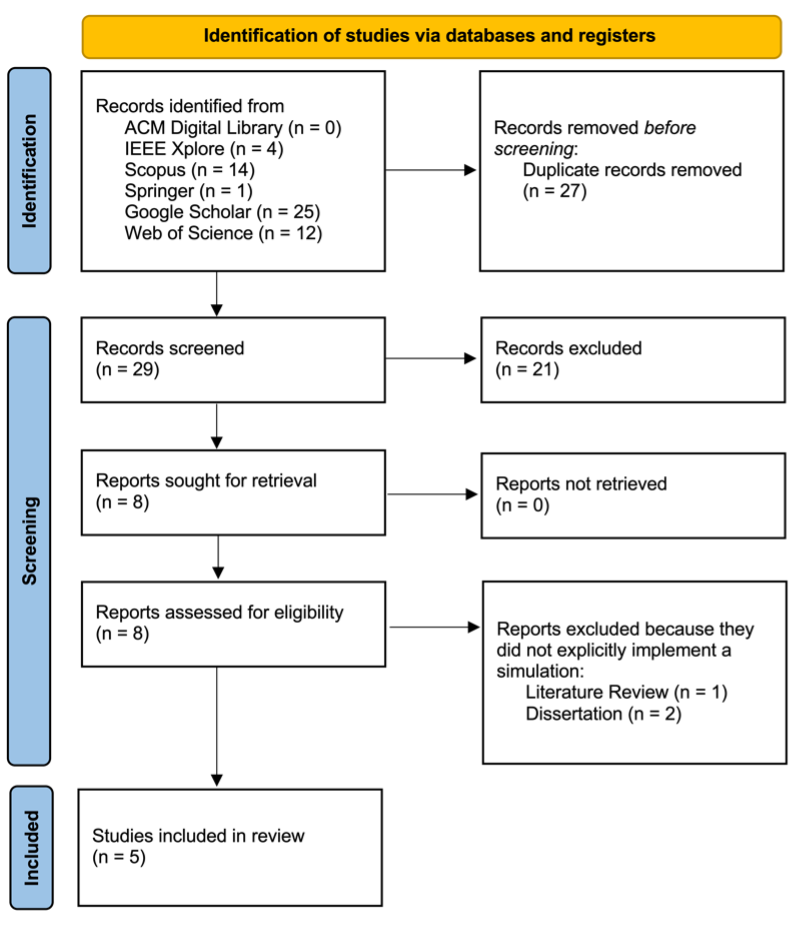}
    \caption{PRISMA 2020 Flow Diagram Illustrating the Study Selection Process.}
    \label{fig:prisma}
\end{figure*}

The study selection followed a systematic, multi-phase process appropriate for a systematic mapping study. A PRISMA (Preferred Reporting Items for Systematic Reviews and Meta-Analyses) 2020~\cite{page2021} flow diagram is included to ensure transparency across the identification, screening, eligibility, and inclusion stages. In the first phase, a structured search was conducted across six databases, yielding 56 initial results. In the second phase, titles and abstracts were screened for relevance, and articles unrelated to software engineering or simulation (e.g., those addressing physical waterfalls) were excluded. In the third phase, full-text articles were retrieved and assessed against the inclusion and exclusion criteria. Three papers were removed because they did not explicitly simulate the waterfall model (i.e., Maxwell-Sinclair~\cite{maxwell2016}, Feddock~\cite{feddock2016}, Negrete et al.~\cite{negrete2023}). In the final phase, duplicate entries resulting from cross-database overlaps were identified and eliminated. The final selection consisted of five unique studies that directly addressed simulation of the waterfall model. These studies formed the basis for data extraction and synthesis. Table~\ref{tab:papers-by-db} presents the distribution of retrieved and retained articles by source. The study selection process is also summarized in the PRISMA flow diagram (Figure~\ref{fig:prisma}), which illustrates the identification, screening, eligibility, and inclusion stages of the mapping study. Table~\ref{tab:selected-studies} summarizes the selected studies, including authors, publication year, title, and reference.

\begin{table*}[ht]
    \centering
    \caption{Papers Identified by Database}
    \label{tab:papers-by-db}
    \renewcommand{\arraystretch}{1.2}
    \begin{tabular}{lcccc}
        \toprule
        \textbf{Source} & \textbf{Returned} & \textbf{Retained} & \textbf{Irrelevant} & \textbf{Duplicate} \\
        \midrule
        ACM Digital Library & 0 & 0 & 0 & 0 \\
        IEEE Xplore & 4 & 1 & 3 & 0 \\
        Scopus & 14 & 3 & 7 & 4 \\
        Springer & 1 & 0 & 0 & 1 \\
        Google Scholar & 25 & 1 & 13 & 11 \\
        Web of Science & 12 & 0 & 1 & 11 \\
        \bottomrule
    \end{tabular}
\end{table*}

\begin{table*}[htbp]
\caption{Summary of Selected Studies on Simulating the Waterfall Model, Including Authors, Publication Year, Title, and Source}
    \label{tab:selected-studies}
\centering
\renewcommand{\arraystretch}{1.2}
\begin{tabular}{p{1cm} p{4.2cm} p{1.4cm} p{6.4cm} p{1.2cm}}
\toprule
\textbf{ID} & \textbf{Author(s)} & \textbf{Year} & \textbf{Title} & \textbf{Ref.} \\
\midrule
P1 &
Luisanna Cocco, Katiuscia Mannaro, Giulio Concas, and Michele Marchesi &
2011 &
Simulating Kanban and Scrum vs.\ Waterfall with System Dynamics &
[2] \\

P2 &
Youssef Bassil &
2012 &
A Simulation Model for the Waterfall Software Development Life Cycle &
[1] \\

P3 &
Prakriti Trivedi and Ashwani Sharma &
2013 &
A Comparative Study between Iterative Waterfall and Incremental Software Development Life Cycle Model for Optimizing the Resources Using Computer Simulation &
[14] \\

P4 &
Taiga Mitsuyuki, Kazuo Hiekata, Takuya Goto, and Bryan Moser &
2017 &
Evaluation of Project Architecture in Software Development Mixing Waterfall and Agile by Using Process Simulation &
[9] \\

P5 &
Antonios Saravanos and Matthew X.\ Curinga &
2023 &
Simulating the Software Development Lifecycle: The Waterfall Model &
[13] \\
\bottomrule
\end{tabular}
\end{table*}

\subsection{Data Extraction and Synthesis}

Following study selection, a structured data extraction form was used to systematically capture key information from each included paper. The form recorded bibliographic metadata (authors, publication year, title, and venue) as well as details of the simulation approach. These details included the simulation methodology employed (e.g., discrete-event simulation or system dynamics), the software tools or platforms used (e.g., Simphony.NET or SimPy), the application context (e.g., education, project planning), and the study’s main findings and implications. The selection and extraction forms were adapted from systematic literature review templates (Kitchenham et al.~\cite{kitchenham2010}) and expanded with mapping categories (methods, tools, venue, year, geography, fidelity).

In standard practice, a subset of papers would be used to pilot the data extraction process to ensure consistency. However, given the small sample size, the extraction process was instead applied to the entire set of included studies and repeated in full to ensure accuracy. This double-pass approach allowed thorough verification of the extracted data and resolution of discrepancies through cross-checking and reflection. In this context, it effectively replaced the need for a pilot subset, as the limited number of studies made comprehensive validation feasible. 

The extracted data were analyzed using a thematic synthesis approach, which enabled the studies to be categorized and compared based on common themes aligned with the research questions. Thematic synthesis is a qualitative method that involves identifying, organizing, and interpreting patterns or themes across a set of studies. Originally developed in the health and social sciences, it has been increasingly adopted in software engineering to facilitate systematic comparison of qualitative attributes, such as methodologies, tools, applications, outcomes, and challenges. This approach is particularly valuable in mapping studies, where the aim is to organize heterogeneous research that does not lend itself to statistical meta-analysis. Patterns and trends were analyzed across simulation techniques, geographic origins, publication types, and study focus. Where appropriate, results were presented in charts and summary tables to provide a clearer overview of the landscape. This synthesis offered structured insights into current practices, emerging trends, and research gaps in the simulation of the waterfall model.

In this mapping study, thematic synthesis refers to the analytical process of grouping and organizing studies by key themes, while the narrative approach describes how these results are presented and interpreted through descriptive text rather than statistical aggregation. The narrative approach involves synthesizing and explaining the findings of included studies using descriptive summaries and comparisons, rather than pooling results statistically. Consistent with systematic mapping study practice, this approach was used to explore heterogeneity across the mapped studies. Differences in simulation methods, tool selection, geographic and temporal trends, and variations in model fidelity were compared and discussed to highlight sources of variation. Where reported, minimal quality and transparency indicators (e.g., model or tool disclosure, verification or validation notes) were recorded, but study evidence was not graded using systematic literature review schemes (Dybå \& Dingsøyr~\cite{dyba2008}).

This systematic mapping study was not registered in a public protocol database, and no formal mapping protocol was prepared prior to its conduct. 

\begin{table*}[htbp]
\caption{Characteristics of Selected Studies Categorized by Publication Type, Geographic Origin, Simulation Tool, and Methodology}
\label{tab:study_characteristics}
\centering
\renewcommand{\arraystretch}{1.2}
\begin{tabular}{p{1cm} p{3.2cm} p{3.2cm} p{3.2cm} p{3.2cm} p{2.2cm}}
\toprule
\textbf{ID} & \textbf{Publication Type} & \textbf{Country of Origin} & \textbf{Simulation Tool} & \textbf{Methodology} & \textbf{Focus} \\
\midrule
P1 & Conference proceeding & Italy & Not specified & System dynamics & Comparative \\
P2 & Journal article & Lebanon & Simphony.NET & Discrete-event & Waterfall \\
P3 & Conference proceeding & India & Simphony.NET & Discrete-event & Comparative \\
P4 & Journal article & Japan & Custom solution & Discrete-event & Comparative \\
P5 & Journal article & United States & SimPy & Discrete-event & Waterfall \\
\bottomrule
\end{tabular}
\end{table*}

\section{Results and Discussion}

This section presents the findings of the systematic mapping study, organized around the research questions defined in Section~2. Using thematic synthesis, the five included studies were analyzed across several dimensions: simulation methodology, tool usage, geographic origin, publication trends, comparative scope, temporal distribution, and model fidelity. The analysis provides a comprehensive overview of how the waterfall model has been represented in simulation-based research and highlights key patterns, divergences, and areas for future investigation.

\subsection{Simulation Methodologies (RQ1.1)}

Two dominant simulation methodologies were identified across the included studies: discrete-event simulation (DES) and system dynamics (SD). DES was employed in four of the five studies (P2, P3, P4, P5), making it the most frequently used technique. DES models the passage of time as a sequence of discrete events, each representing a change in system state. This approach aligns well with the sequential structure of the waterfall model, in which development progresses through phases such as requirements, design, implementation, testing, and deployment. The DES studies examined factors including resource constraints, task delays, phase dependencies, and productivity rates. For example, Bassil (P2) and Saravanos and Curinga (P5) modeled performance metrics such as project duration and resource utilization, showing how early-phase bottlenecks can cascade into later stages. Mitsuyuki et al.\ (P4) implemented a custom simulator but explicitly evaluated scenarios using DES, organizing work into six task types: planning, design, implementation, unit test, integration test, and review. By contrast, SD was used in one study (P1) to compare waterfall with Kanban and Scrum, emphasizing feedback structures and system-level dynamics over time.

\begin{figure*}[ht]
    \centering
    \includegraphics[width=1.0\linewidth]{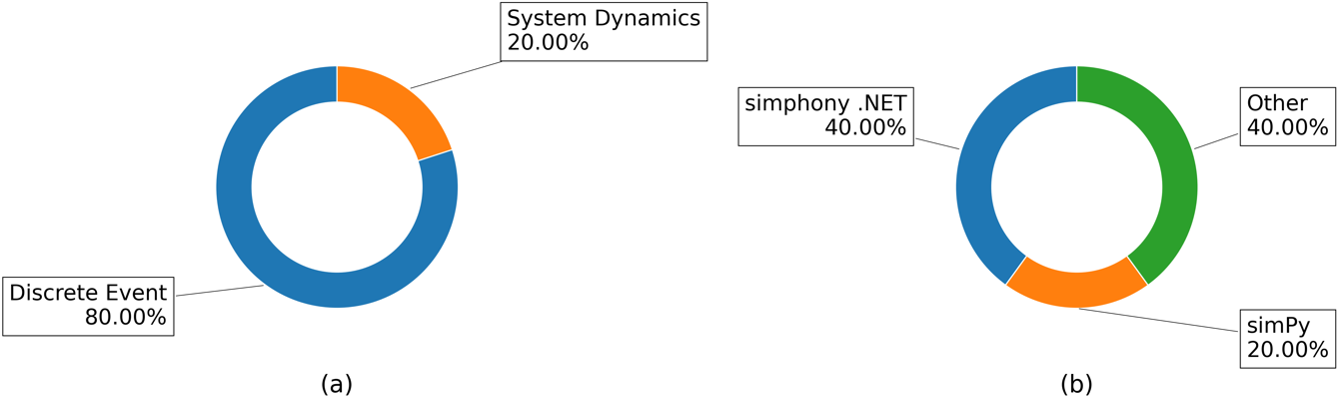}
    \caption{Pie Charts Illustrating (a) the Type of Approach Used for the Simulation and (b) the Tool That was Used.}
    \label{fig:approach-tool}
\end{figure*}

In summary, DES appears well-suited for operational analyses of waterfall workflows and was employed in four of the five studies, while SD provided a single systems-level comparative perspective. The distribution of methodologies is shown in Figure~\ref{fig:approach-tool}(a) and corresponds with the characterization in Table~\ref{tab:study_characteristics}.

\subsection{Simulation Tools (RQ1.2)}

The simulation tools used in the reviewed studies varied in sophistication and accessibility, reflecting broader trends in software process modeling. Simphony.NET was the most frequently used tool, appearing in two studies (P2, P3). Developed at the University of Alberta, Simphony.NET is a general-purpose, Windows-based discrete-event simulation platform originally designed for construction and process management. Its drag-and-drop modeling interface and event-driven architecture make it well suited for representing sequential workflows such as the waterfall model. Its repeated use suggests both robustness and adaptability to software-engineering contexts, despite its origins in construction simulation.

By contrast, SimPy, a Python-based simulation library, was used in one study (P5). SimPy supports event-based simulations through process generators and has gained popularity due to its integration with Python’s extensive ecosystem, ease of use, and transparency. In this study, SimPy was applied to simulate the full lifecycle of software development projects under the waterfall framework, including task durations, resource allocation, and interdependencies. The adoption of SimPy in this more recent work suggests a trend toward open-source, flexible, and accessible simulation environments, particularly in educational and research settings.

Two studies (P1, P4) did not specify their simulation tools, which limits both reproducibility and transparency. While P1 employed system dynamics (likely using tools such as Vensim or Stella), the omission prevents definitive conclusions. Overall, the range of tools suggests that although Simphony.NET has historically been favored, the adoption of Python-based solutions like SimPy may indicate a shift toward more lightweight and customizable simulation frameworks in software-engineering research. The distribution of tool usage is shown in Figure~\ref{fig:approach-tool}(b). 

\subsection{Geographic Distribution of Research (RQ2.1)}

\begin{figure*}
    \centering
    \includegraphics[width=1.0\linewidth]{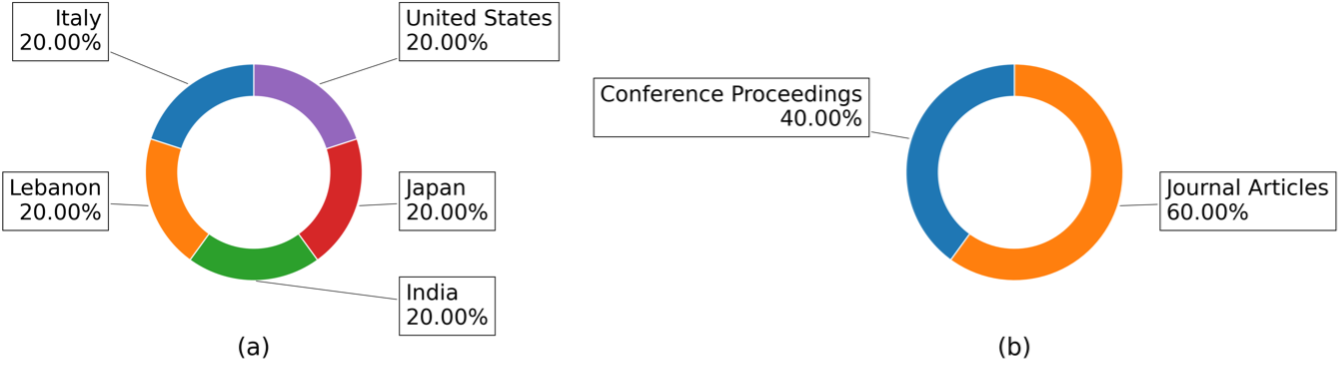}
    \caption{Pie Charts Illustrating (a) the Country of Origin of the Authors of the Paper and (b) Whether the Publications Appeared as Conference Proceedings or Journal Articles.}
    \label{fig:geo-pubtype}
\end{figure*}

The studies originate from five different countries, reflecting the global interest in simulating the waterfall model: Italy (P1), Lebanon (P2), India (P3), Japan (P4), and the United States (P5). Contributions from both Western and Eastern institutions suggest that the waterfall model continues to attract academic attention across diverse cultural and technological environments, even as agile and hybrid methodologies dominate industrial practice. At the same time, the low number of studies from each region indicates that simulation of the waterfall model remains a relatively underexplored niche, potentially limited by factors such as tool accessibility, research funding, or curricular emphasis. The breakdown is illustrated in Figure~\ref{fig:geo-pubtype}(a).

\subsection{Publication Types (RQ3.1)}

Research on simulating the waterfall model has appeared in both journals and conference proceedings. Three of the five studies (P2, P4, P5) were published in peer-reviewed journals, while the remaining two (P1, P3) were presented at conferences. This distribution reflects a broader trend in software-engineering research, where early or exploratory work is often introduced at conferences, while more mature and in-depth studies are published in journals. The earlier studies (2011--2013) were primarily conference papers (with the exception of P2), whereas the more recent works (2017--2023) were journal articles (P4, P5). This shift suggests both a maturation of scholarly interest in simulating traditional process models and improvements in methodological rigor and reporting. Journal publication may also signal stronger institutional or funding support for simulation-based research and growing recognition of simulation methods as valid tools in empirical software engineering. The distribution is shown in Figure~\ref{fig:geo-pubtype}(b).

\subsection{Comparative Scope of Simulations (RQ3.2)}

Three studies (P1, P3, P4) conducted comparative simulations that examined the waterfall model alongside alternative development methodologies, including Scrum, Kanban, iterative, and incremental models. These studies generally sought to evaluate how the waterfall model performs under different conditions relative to more adaptive or iterative approaches. For example, Trivedi and Sharma (P3) contrasted resource optimization between iterative waterfall and incremental models, while Mitsuyuki et al.\ (P4) simulated hybrid scenarios that combined waterfall and agile elements. Such comparative simulations highlight trade-offs between predictability and flexibility, particularly in contexts such as education or regulatory compliance, where linear models may still retain value. By contrast, only two studies (P2, P5) focused exclusively on simulating the waterfall model in isolation. These works analyzed waterfall processes in greater detail, addressing aspects such as task dependencies, resource usage, and failure propagation. The predominance of comparative studies reflects both the need to identify context-appropriate development models and the contemporary relevance of hybrid strategies. This trend suggests that simulation is being used not merely to analyze the waterfall model in isolation, but to assess its utility and limitations in relation to evolving methodologies.

\subsection{Temporal Distribution of Studies (RQ3.3)}

The temporal analysis shows that the selected studies were published between 2011 and 2023. The earliest identified work (P1) appeared in 2011, while the most recent (P5) was published in 2023. Although the overall number is small---only five studies across more than two decades---the relatively even distribution suggests a sustained, if limited, scholarly interest in simulating the waterfall model. This continuity indicates that, while not a dominant research theme, the topic has persisted in academic discourse.

The scarcity of relevant studies is itself a noteworthy finding. Despite the foundational role of the waterfall model in software-engineering history and its continued presence in educational curricula, regulatory frameworks, and documentation practices, few publications have used simulation to model or analyze it in isolation. This raises questions about research priorities and possible blind spots in the literature. The decline in industrial use of the waterfall model may partly explain the reduced academic focus, especially as agile, DevOps, and hybrid methodologies have gained prominence. Researchers may also perceive traditional, sequential models as offering limited novelty compared with the dynamic features of iterative and adaptive lifecycles.

Terminology and framing may present another explanation. Some simulation studies may have modeled plan-driven processes without explicitly labeling them ``waterfall'', particularly in comparative contexts. As a result, relevant work may have been excluded by the search strategy, even if the underlying processes were conceptually similar.

Nonetheless, the appearance of recent studies, especially those using modern tools such as SimPy, points to a shift in emphasis, from industrial process modeling toward educational and exploratory uses of simulation. These newer works often prioritize transparency, accessibility, and instructional value. This trend suggests growing interest in using simulation both to teach foundational concepts in software development and to benchmark traditional models against contemporary ones under controlled conditions.

\begin{figure*}
    \centering
    \includegraphics[width=0.8\linewidth]{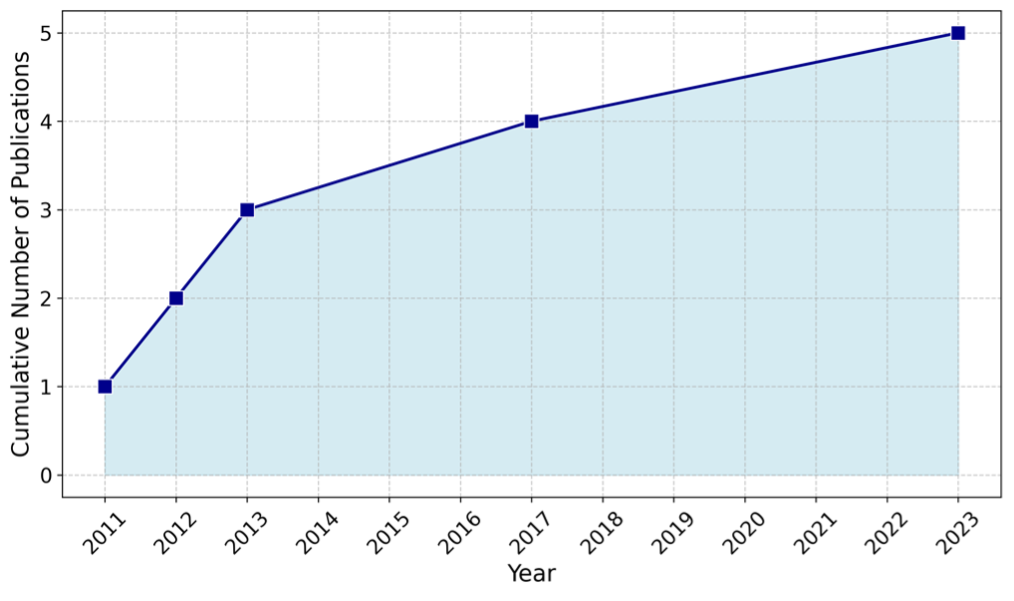}
    \caption{The Cumulative Number of Publications by Year.}
    \label{fig:cumulative-pubs}
\end{figure*}

In sum, the limited number of simulation-focused studies on the waterfall model should be viewed not only as a constraint but also as an opportunity. The field remains underexplored, leaving room for future research that reconsiders the waterfall model’s role, not merely as a legacy methodology but also as a tool for pedagogy, process experimentation, and comparative analysis in hybrid software development environments. The cumulative number of publications by year is shown in Figure~\ref{fig:cumulative-pubs}.

\subsection{Model Fidelity (RQ4.1)}

Fidelity to Royce’s waterfall model can be assessed along two dimensions, given the three versions presented in his original paper: phase granularity (the extent to which the simulated phases match Royce’s seven) and control flow (the policy governing movement between phases---single pass with no feedback, two passes through the lifecycle, or a single pass with bounded backflow to the immediately prior phase). This analysis excludes the more innovative models also introduced by Royce. All five studies reviewed reduced phase granularity, none retained all seven phases, but they differed in control flow. P1 described waterfall as a prescriptive sequence in which each phase must be completed before the next begins, and in which changed requirements were not revisited until an entire cycle had finished. Rework was detected at the end of the iteration and added delay, rather than enabling an immediate return to a previous phase. By contrast, P2, P3, P4, and P5 made backflow explicit: after a task was completed, an error branch could send work back to the immediately preceding phase.

Taken together, these adaptations are not incidental; they signal how the waterfall model is interpreted and applied in simulation research. Consolidated phases reduce model complexity and align with common teaching and tooling constraints, while feedback paths acknowledge routine rework (e.g., requirement changes, defect discovery) and the prevalence of hybrid practices in contemporary contexts. The absence of a faithful seven-phase implementation supports long-standing critiques that Royce’s original, single-pass depiction is rare in practice and often counter-normative in both education and industry. For future work, we encourage authors to report fidelity explicitly. For example: (1) which Royce phases are merged or omitted; (2) what backflow is permitted and under what conditions; and (3) the criteria used to declare phase completion. Such disclosure would improve comparability across studies and support reproducibility without precluding pragmatic adaptations to tools or instructional aims.

\section{Threats to Validity}

Our discussion of limitations follows software-engineering review conventions (adapted from Kitchenham et al.~\cite{kitchenham2010}) but is scoped to mapping outcomes rather than effect synthesis. As with all systematic mapping studies, this work is subject to several potential limitations, categorized using common dimensions in software-engineering research: construct, internal, and external validity. 

\subsection{Construct Validity}

Construct validity concerns whether our operationalization of the topic, the databases searched, the time window, and especially the title-only query (``simulation'' OR ``simulating'') AND ``waterfall'', actually captured the body of work that explicitly simulates the waterfall model. Although this search string was chosen to maximize recall while keeping screening manageable, it risks yielding both false positives and false negatives. False positives are items that were retrieved but not actually relevant, for example: (1) articles using ``waterfall'' in unrelated domains (e.g., geological waterfalls, financial ``waterfall charts'', ``waterfall effect'' in physics) that also mentioned simulation; (2) works that mentioned the waterfall model but do not conduct a simulation (e.g., conceptual critiques, surveys, teaching notes); and (3) studies in which ``waterfall'' is only a contextual reference within a broader simulation (e.g., hybrid processes) without explicitly simulating the waterfall process itself. False negatives are work that was relevant but not retrieved, for example: (1) studies that simulated a plan-driven or linear SDLC but avoided the word ``waterfall'', instead using terms such as ``sequential SDLC'', ``phase-gate'', ``stage-gate'', or ``traditional model''; (2) studies that ran simulations but whose titles used other labels (e.g., process modeling, discrete-event analysis, system dynamics study) without the words ``simulation'' or ``simulating''; and (3) studies mentioning ``waterfall'' only in the abstract or body text (excluded because our search was title-only). In short, our query traded some precision (increasing the risk of false positives) for recall (while still risking false negatives due to title phrasing). We mitigated this risk by searching six major databases and manually screening results, but we acknowledge that relevant work may exist under alternative terms or labels. Future replications could reduce these risks by expanding to abstract and keyword searches, incorporating controlled vocabulary (e.g., plan-driven, phase-gate), and conducting both backward and forward citation chasing.

\subsection{Internal Validity}

Internal validity concerns the consistency with which studies were screened and data extracted. Because a single reviewer conducted both phases, judgments requiring interpretation, such as whether a model truly ``simulates'' waterfall, distinguishing discrete-event simulation from system dynamics, or assessing fidelity to Royce, may reflect individual bias. The small corpus (five studies) made a complete double-pass feasible. Detailed notes and extraction logs were maintained to ensure consistency: each inclusion or exclusion decision and each coded attribute (method, tool, venue, geography, fidelity, comparative scope) was rechecked at least twice. Even so, without independent reviewers we did not compute inter-rater agreement, and errors cannot be fully ruled out, such as misclassifying a hybrid as ``waterfall-only'', inconsistently handling missing tool information (``not specified'' versus ``unknown''), or selection drift over time. These limitations mean the mapping should be interpreted as a careful but single-analyst synthesis rather than a consensus view. Future replications could strengthen internal validity by involving at least two independent reviewers for screening and extraction, running a calibration round on a subset, reporting inter-rater reliability (e.g., Cohen’s kappa), preregistering a protocol, piloting the extraction form, and publishing an auditable replication package (including search strings, screening decisions, and coded data).

\subsection{External Validity}

External validity addresses the extent to which the findings can be generalized beyond the sample. With only five studies meeting the inclusion criteria over a 24-year span, the patterns reported should be read as indicative rather than definitive. The sample is small, heterogeneous (different methods, tools, and model variants), and unevenly distributed across venues and countries, which limits the generalizability of observed trends. Our scope further narrows generalizability by design. We restricted inclusion to peer-reviewed, English-language publications and used a title-only search, meaning results best reflect the subset of work that labels itself explicitly as ``simulation'' and ``waterfall'' in the title. Relevant studies in other languages, in grey literature (e.g., industry reports), or under alternative labels for plan-driven processes may be underrepresented. Moreover, several included papers omitted tool details or provided limited modeling disclosures, reducing comparability and making it difficult to infer field-wide practices. In short, this mapping provides a cautious snapshot of a niche research area: papers that explicitly present waterfall simulations in peer-reviewed, English-language venues between 2000 and 2024. Claims beyond that population, such as about industry practice, non-English scholarship, or plan-driven simulations that do not use the term ``waterfall'' in the title, should be made with care. Future replications could strengthen external validity by broadening searches to titles, abstracts, and keywords; incorporating synonyms for plan-driven lifecycles; including non-English sources and selected grey literature; and using citation chasing to surface studies outside database indexing or title phrasing.

\subsection{Conclusion Validity}

Conclusion validity concerns whether the conclusions drawn are warranted by the evidence collected. Because the corpus is small and heterogeneous, our synthesis is qualitative and descriptive. We did not conduct statistical tests or a meta-analysis; therefore, our statements reflect patterns observed within this sample rather than predictions about waterfall simulations in general. Consistent with the aims of a mapping study, we did not grade the strength of evidence for individual papers or pool effect estimates. In line with guidance on evidence assessment in software-engineering reviews (cf.\ Dybå \& Dingsøyr~\cite{dyba2008}), we treat the findings as descriptive signposts that organize reported methods, tools, and contexts. We did not apply a formal risk-of-bias checklist. Instead, we relied on practical transparency signals, such as whether a study specified its tool, described its model, noted any verification or validation, and explained how closely it followed Royce’s formulation. These signals help readers judge credibility but are not substitutes for a structured appraisal. The implication is that our mapping reflects what this small set of papers reports, not what necessarily holds across the broader literature or in practice. Strong causal or performance claims should not be drawn from these results. Future replications could strengthen conclusion validity by adopting or designing bias and quality assessment tools tailored to simulation in software engineering, preregistering a protocol, piloting and publishing the extraction form, reporting inter-rater agreement when multiple reviewers are involved, and releasing a replication package with search strings, screening decisions, coded data, and, where possible, model code.

\section{Conclusions, Limitations, and Future Work}

This paper presents, to our knowledge, the first systematic mapping study dedicated to how the waterfall model has been simulated in the software-engineering literature. Although the waterfall model is widely taught and historically influential, its explicit simulation has received relatively little direct scholarly attention. Of the 56 studies initially retrieved through structured database searches, only five met the inclusion criteria for simulating the waterfall model in a clear and documented way. These five studies were examined in detail to identify methodological patterns, tool usage, geographic distribution, publication trends, comparative focus, and adherence to Royce’s original formulation.

The mapping revealed that discrete-event simulation (DES) is the dominant approach for modeling the waterfall process. Four of the five studies employed DES, while one used system dynamics (SD). Tool usage was limited and unevenly reported: two studies used Simphony.NET, one used SimPy, and two did not specify the tool, reducing transparency and reproducibility. The studies originated from five countries: Italy, Lebanon, India, Japan, and the United States. Publication venues included both journals and conference proceedings, with three studies appearing in journals and two in conferences. Three studies employed comparative designs, simulating the waterfall model alongside other approaches such as agile, iterative, or hybrid methods, while two modeled the waterfall process alone.

Despite originating from different contexts, all five studies diverged from Royce’s original specification of the waterfall model. None preserved all seven phases or employed a strictly single-pass formulation with no feedback. Instead, most allowed limited backflow to earlier stages or adopted simplified, loop-based structures. This divergence reflects the practical reality that pure waterfall processes are rarely implemented without adaptation. Phase consolidation and feedback loops likely reflect both the constraints of simulation tools and the needs of educational or comparative analyses.

The findings carry several implications for researchers, educators, and practitioners. For researchers, simulation provides a structured and reproducible means of exploring software process behavior. The waterfall model, with its deterministic and phase-based structure, remains useful for such exploration, particularly in comparative contexts. Lightweight and transparent simulation environments such as SimPy are especially effective for demonstrating the effects of early-stage decisions, resource allocation, and process rigidity. For practitioners, particularly in safety-critical or regulated domains, simulations of the waterfall process can support planning, risk modeling, and resource management.

This study also has limitations. The search strategy was restricted to title-only queries using the terms ``simulation'' or ``simulating'' and ``waterfall'', and it was limited to English-language, peer-reviewed publications between 2000 and 2024. Although this scope ensured that included studies explicitly addressed the topic, it may have omitted relevant work using different terminology or labels. The number of included studies is small, and findings are therefore descriptive rather than statistically generalizable. Additionally, all screening and data extraction were conducted by a single reviewer, though each decision was double-checked during a second pass. No formal inter-rater reliability was calculated, and no meta-analysis was performed. These limitations mean that conclusions should be interpreted cautiously and viewed as indicative rather than definitive.

Several opportunities for future research arise from this work. Future mapping studies should broaden their scope to include abstracts and keywords, incorporate synonyms for plan-driven or phase-based processes, and examine gray literature and non-English sources. Authors simulating the waterfall model should adopt consistent and transparent reporting practices, including details on phase structure, control-flow policies, modeling assumptions, tool usage, and the availability of code and data. Open-source platforms such as SimPy offer a promising basis for reusable simulation models that can serve as reference implementations for both research and instruction. Publishing simulation code, parameter files, and experimental outputs would further enhance transparency and enable replication. Researchers are also encouraged to explore hybrid models that integrate waterfall phases with iterative or agile substructures, as these increasingly reflect contemporary practice.

Far from being an artifact of the past, the waterfall model, when paired with robust simulation, remains a relevant, evidence-based framework for planning and managing projects where stability, compliance, and disciplined execution are essential. Its continued use in domains such as aerospace, defense, and medical software underscores its importance in contexts where predictability, traceability, and compliance are critical. Simulation, when implemented with methodological rigor and transparency, provides a valuable means of studying the behavior of such processes under varying conditions. It enables controlled experimentation with process variables, supports the evaluation of planning assumptions, and highlights trade-offs between structure and flexibility. Rather than being viewed solely as obsolete, the waterfall model, when combined with simulation, offers a meaningful and enduring framework for analyzing software development workflows.

\end{document}